# Dynamic Buffer Management for Multimedia QoS in Beyond 3G Wireless Networks

Suleiman Y. Yerima and Khalid Al-Begain

*Abstract*— This paper investigates a dynamic buffer management scheme for QoS control of multimedia services in beyond 3G wireless systems. The scheme is studied in the context of the state-of-the-art 3.5G system i.e. the High Speed Downlink Packet Access (HSDPA) which enhances 3G UMTS to support high-speed packet switched services. Unlike earlier systems, UMTS-evolved systems from HSDPA and beyond incorporate mechanisms such as packet scheduling and HARQ in the base station necessitating data buffering at the air interface. This introduces a potential bottleneck to end-to-end communication. Hence, buffer management at the air interface is crucial for end-to-end QoS support of multimedia services with multiplexed parallel diverse flows such as video and data in the same end-user session. The dynamic buffer management scheme for HSDPA multimedia sessions with aggregated real-time and non real-time flows is investigated via extensive HSDPA simulations. The impact of the scheme on end-to-end traffic performance is evaluated with an example multimedia session comprising a real-time streaming flow concurrent with TCP-based non real-time flow. Results demonstrate that the scheme can guarantee the end-to-end QoS of the real-time streaming flow, whilst simultaneously protecting the non real-time flow from starvation resulting in improved end-to-end throughput performance.

## I. INTRODUCTION

Despite the penetration of 3G UMTS mobile communication systems worldwide, increasing demand for ubiquitous wireless access has continued to drive advancements to support mobile broadband services. The 3rd Generation Partnership Project (3GPP) is currently developing the Long Term Evolution (LTE) technologies aimed at getting the 3G infrastructure to provide data rates up to 100 Mbps. An important step in this direction has been the introduction of High-Speed Downlink Packet Access (HSDPA) [1-4]. HSDPA is a 3.5G wireless system standardized as a set of technological advancements to UMTS in order to improve network capacity and increase the peak data rates to 14.4 Mbps for downlink packet traffic. The Release 7 version of HSDPA is expected to push the downlink data rate to 28 Mbps.

HSDPA utilizes a common downlink shared channel known as high speed downlink shared channel (HS-DSCH) and employs fast link adaptation for downlink data transfer to mobile stations. The fast link adaptation feature is based on Adaptive Modulation and Coding (AMC), Hybrid Automatic Repeat reQuest (HARQ) and a shorter minimum allocation time (transmission time interval, TTI) of 2ms. In addition to these physical layer features, the packet scheduling functionality is moved from the centralized radio network controller (RNC) to the base station (Node B), where it is embedded in a new MAC entity known as MAC-hs.

The higher data rate capability of HSDPA will allow application developers to create content rich 'multimedia' applications and services, typically consisting of a number of classes of media or data- with different Quality of Service (QoS) requirements- being concurrently downloaded to a single user [5]. Additionally, support for multimedia services/applications with different classes of media/flows is a key requirement of the UMTS-HSDPA system [1], [3],[6]. Furthermore, packet scheduling and HARQ mechanisms in the Node B necessitate buffering at the edge of the air interface therefore posing a potential bottleneck to end-to-end multimedia traffic. In [3], the necessity for Node B buffer management in general, to improve traffic performance has been emphasized but without proffering solutions.

The study in this paper is motivated by the aforementioned; hence, a proposed buffer management scheme for QoS control of end-user HSDPA multimedia traffic with concurrent real-time and non real-time flows such as streaming video and data (in a single user session), is presented and evaluated. The scheme, termed the *dynamic time-space priority* (D-TSP) buffer management incorporates time-priorities and space-priorities as well as dynamic transmission priority switching between the aggregated flows to suit changing QoS requirements.

Time-priorities and space-priorities are mechanisms used for traffic class differentiation in buffer/queue management schemes. Time-priorities provide the ability to control the process that outputs data from the buffers, thus giving preferential delay treatment to one/some class(es) of traffic over the other(s) in order to control delay and jitter. Space-priorities offer preferential loss treatment to one/some class(es) of traffic over others by controlling the buffer access mechanism. According to [7], buffer management is a fundamental technology to provide QoS control mechanisms, by controlling the assignment of buffer resources among different flows or aggregation of flows according to certain policies. Hence, most buffer management schemes in the literature focus on space-priorities and fall into three main categories from resource management viewpoint [8],[9]: complete buffer partitioning (CBP), complete buffer sharing

Manuscript received 27 September, 2009.
  Suleiman Y. Yerima is with the Integrated Communications Research Centre (ICRC), Faculty of Advanced Technology, University of Glamorgan, Pontypridd, Mid Glamorgan, CF37 1DL, United Kingdom (Phone: +44 (0)1443 48 3612; e-mail: syerima@glam.ac.uk).
  Khalid Al-Begain, is the Head of Integrated Communications Research Centre (ICRC), University of Glamorgan (e-mail: kbegain@glam.ac.uk).

(CBS), and partial buffer sharing (PBS).

CBP policy segments a given buffer space into multiple dedicated queues according to the differentiated classes of traffic. The allocation can be static or dynamic. CBS on the other hand, admits traffic of all classes as long as space is available in the buffer. In order to overcome the lack of service differentiation in CBS, the *pushout* policy allows *low priority* packets to be discarded by an incoming *higher priority* packet. PBS is a widely studied buffer management policy that provides preferential access to buffer space to high priority packets at the expense of loss of lower priority packets using a threshold. When exceeded, the threshold bars low priority packets from admission into the buffer while high priority packets continue to access the buffer.

Buffer management schemes based on the above mentioned policies have been proposed for wireless networks to facilitate effective sharing of bottleneck resources for enhanced QoS support. For example, [10] investigates buffer management within the HSDPA Radio Access Network to alleviate the impact of unreliable and time-varying link on non-interactive video streaming services. The paper proposes proactive buffer management schemes with data differentiation to improve MPEG-4 video quality. The schemes were PBS-based *proactive B-dropping* where the buffer proactively drops packets containing data of B-frames if the buffer occupancy exceeds a given threshold. Similarly,[11] considers several space-priority CBS-based schemes for the Radio Link Control (RLC) layer buffer management for video streaming over a HSDPA shared channels. The schemes are: *Drop New Arrivals* (DNA) which is equivalent to drop-tail CBS; *Drop Random Packet* (DRP) equivalent to *pushout* with *random* replacement; and the *Drop HOL Packet* (DHP) equivalent to *pushout* with *first-in-first-drop* replacement.

In [12], a buffer management scheme is proposed for 3G UMTS networks. The scheme is based on multi-level priority and CBS policy for all buffers at the border and inside the wireless network. Other works related to buffer management for QoS enhancement of traffic in wireless networks can be found in [13-16]. This highlights the growing importance of buffer management-based solutions to improve QoS performance in wireless networks. Nevertheless, QoS management of multimedia sessions with multiplexed services comprising concurrent diverse classes of flows in the user session has largely remained unaddressed.

Existing wireless buffer management solutions being mainly space-priority based cannot adequately provide the joint QoS control required by multimedia traffic with aggregated real-time and non-real-time streams. Hence, D-TSP with its hybrid (time and space) prioritization properties as well as dynamic time priority switching between RT and NRT flows is designed to overcome this problem. In D-TSP, *time priority* allows the delay-sensitive and loss-tolerant real-time (RT) flow to fulfill delay and jitter requirements, whilst *space priority* given to the loss-sensitive but delay tolerant non real-time (NRT) flow enables loss minimization. The dynamic (time) priority switching allows any residual delay tolerance of the real-time flow to be exploited in order to prevent potential (bandwidth) starvation of the parallel non real-time stream. This is particularly crucial at the bottleneck air interface. Hence, D-TSP concept is important not only for downlink multimedia traffic QoS control in HSDPA, but also for other beyond 3G wireless systems with queuing at the air interface.

In this paper, D-TSP is evaluated using HSDPA system simulations and the impact of the scheme on RT streaming flow and concurrent NRT TCP-based data flow in an end-user multimedia session is studied under various HSDPA channel loads. The rest of the paper is organized as follows. Section II discuses buffer management in HSDPA radio access network and gives a description of the D-TSP scheme. Next, HSDPA simulation model for D-TSP performance evaluation is explained; followed by numerical results and analyses. Finally, concluding remarks are given in the last section.

## II. BUFFER MANAGEMENT IN HSDPA RADIO ACCESS NETWORK

In HSDPA, the packet scheduling functionality is performed in the Node B with a transmission time interval (TTI) of 2ms. The basic function of the Node B packet scheduler is to determine which user will receive transmission in the next TTI. The size of the data block transmitted in the TTI frame is determined by the AMC scheme selected, which in turn is based on the channel quality reported via the feedback uplink control channel. In addition to the user equipment (UE) channel quality information, ACK/NACK feedback is also carried on the return uplink channel to enable retransmissions during the HARQ operation.

In the 3GPP HSDPA standards (TS 25.321 MAC specification) [17], packet scheduling is specified as a MAC-hs functionality as shown in Figure 1. Additionally, *priority handling* and *priority queue distribution* functionalities are defined to cater for multiple queues associated with a single user that maintains several flows in the same HSDPA session. Most existing HSDPA packet scheduling algorithms are designed for inter user transmission scheduling and do not address inter-class/inter-flow prioritization for end-users with multiple flows; i.e. the packet scheduling algorithms assume a single flow per user. Where multiple flows or media with different QoS requirements, such as RT voice/video and NRT data exists for a given user, an efficient buffer management scheme can enable inter-class/inter-flow prioritization for each 'multimedia' user, whilst the packet scheduling algorithm provides the per TTI scheduling between the users.

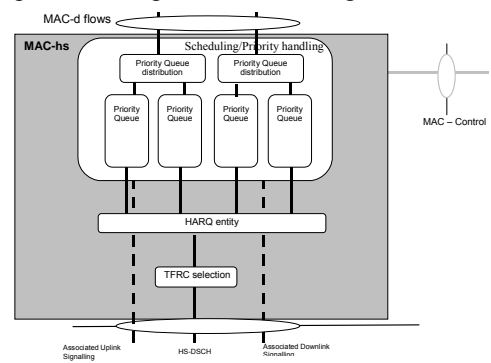

**Figure 1 : HSDPA Node B MAC-hs architecture [17].**

The scheduler schedules MAC-hs SDUs (generated from one or more queued MAC-d PDUs arriving from the RLC)

based on information from upper layers. One UE may be associated with one or more MAC-d flows. Each MAC-d flow contains HS-DSCH MAC-d PDUs for one or more priority queues. Hence, in our approach, the buffer management algorithm (BMA) can be considered a sub-function of the MAC-hs scheduler which oversees the responsibility of handling priorities for multiple flows (priority queues) associated with the same UE in the MAC-hs. With this approach, classical packets scheduling algorithms such as Round Robin, Max throughput, Max-C/I, proportional fair, M-LWDF, etc. can be applied for inter-user transmission scheduling while the BMA handles inter-class transmission prioritization between the RT and NRT flows (through its time priority policy).

The BMA performs buffer admission control (BAC) on arriving MAC-d PDUs according to the priority scheme's BAC mechanism. The BMA also determines the MAC-d flow (identifies the UE priority Queue ID) from which the MAC-d PDUs will be assembled into MAC-hs SDUs for transmission to the UE on a given HARQ entity. If priority switching is enabled (as in the case of a dynamic buffer management scheme like D-TSP proposed in this paper), the priority switching algorithm is consulted to identify the correct MAC-d flow (Queue ID) for next transmission. The scheduler then schedules the MAC-d flow on a selected HARQ process, if the scheduling algorithm has allocated the next transmission opportunity to the UE associated with the MAC-d flow.

Thus, the buffer management algorithm as a sub-entity of the MAC-hs scheduler, is executed on each of the UE's MAC-d flows i.e. the priority handling for UE's with multiple flows per session (i.e. multiplexed RT and NRT flows concurrent in the multimedia connection). The entire scheduling process for multimedia sessions therefore, can be viewed as occurring in a logical hierarchy with the BMA performing buffer admission control and inter-class (inter-flow) scheduling, while the MAC-hs scheduling performs the inter-UE scheduling according to any of the well known scheduling disciplines like Max-C/I, Proportional Fair, etc. This logical scheduling hierarchy is show in Figure 2.

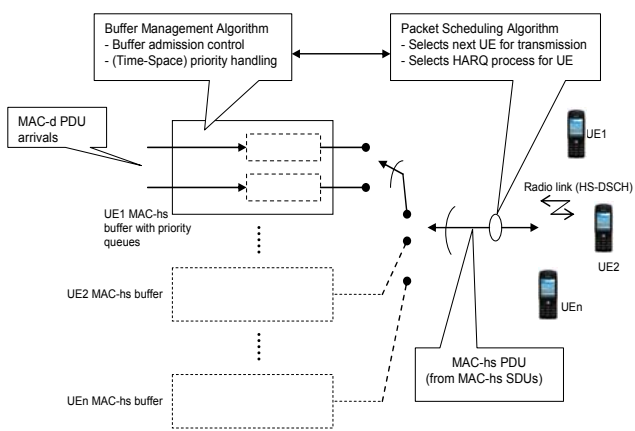

**Figure 2:** Logical hierarchy of $UE_1$ multiplexed RT and NRT MAC-d flow scheduling in HSDPA Node B with intra-user buffer management and inter-user packet scheduling.

*A. MAC-hs PDU transmission at the air interface*

A MAC-hs PDU for High-Speed Downlink Shared Channel (HS-DSCH), as shown in Figure 3, consists of one MAC-hs header and one or more MAC-hs SDUs where each MAC-hs SDU equals one MAC-d PDU [17]. According to 3GPP HSDPA specifications, a maximum of one MAC-hs PDU can be transmitted in a TTI per UE. The MAC-hs header is of variable size. The MAC-hs SDUs in one TTI belong to the same reordering queue in the UE side. This implies that with MAC-hs, the RT and NRT SDUs from a UE's priority queues cannot be multiplexed in one TTI. Thus, when a UE with multiplexed RT and NRT flows is selected for transmission by the MAC-hs packet scheduling algorithm, the BMA policy would determine which priority queue will supply the MAC-hs SDUs (queued MAC-d PDUs) to construct the MAC-hs PDU for transmission on a selected HARQ process. Note that the total size of the MAC-hs PDU is determined by the UE *instantaneous radio conditions* reported in the *CQI* via the uplink (HS-DPCCH) channel. From the CQI, the Adaptive Modulation and Coding (AMC) Scheme is selected along with the number of codes to be used, and this basically determines the size of the MAC-hs PDU to be transmitted to the UE in the TTI.

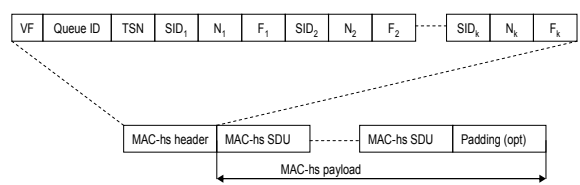

**Figure 3:** MAC-hs PDU structure [17]

Thus, D-TSP is designed to fulfill the BMA roles of priority handling and queue distribution in the Node B MAC-hs for each user with aggregated RT and NRT flows, whilst also allowing optimized QoS control at the air interface in order to enhance end-to-end traffic performance. The D-TSP buffer management scheme is described next.

*B. The D-TSP Buffer Management Scheme*

The basic concept of dynamic time-space priority (D-TSP) is to simultaneously provide transmission (time) priority for the RT flow, and space priority to the NRT data flow of the same end-user. D-TSP utilizes a trade-off mechanism to switch transmission priority to the NRT flow at the expense of slight degradation of RT flow QoS (delay and loss) within the allowable RT QoS constraints. The idea behind D-TSP is to prevent potential NRT flow starvation at the bottleneck i.e. radio interface, without violating the RT flow QoS requirements. The scheme is illustrated in Figure 4. D-TSP is based on a novel Time-Space Priority queuing concept [18], [19], where RT flow and NRT flow destined for the same user are queued using a hybrid priority queuing mechanism.

The RT flow packets, say from a conversational class voice, or real-time streaming video/audio are queued ahead of the NRT flow packets of the same user, for priority scheduling/transmission on the shared channel (i.e. time priority). At the same time, the NRT flow packets, say from background class like (TCP-based) FTP traffic, get space priority in the user's buffer queue because of their loss sensitivity, and lower transmission priority due to their delay tolerance. TSP queuing uses a threshold R (see Fig. 4), to restrict the maximum number of queued RT packets, whilst the NRT flow has unrestricted access to the entire buffer space (i.e. space priority). Threshold R allows loss tolerance of RT flow to be exploited to minimize NRT packet loss.

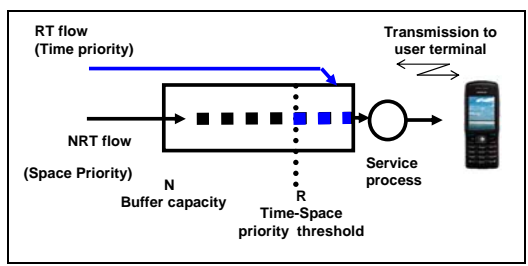

**Figure 4: Basic TSP queuing concept for single user multimedia RT and NRT traffic.**

Since RT flows typically do not employ retransmission protocols, RT packet losses within QoS bounds does not affect higher layer protocol performance. On the other hand, NRT packet losses being typically recovered with higher layer RLC and TCP retransmission protocols adversely affects their performance, resulting in end-to-end throughput degradation.

In [19], we have shown TSP to be an effective queuing mechanism for joint RT and NRT QoS control compared to conventional priority queuing schemes. However, according to 3GPP standards [20],[21], the MAC-hs can incorporate flow control algorithms (Iub flow control) to regulate RNC to Node B data transfer to prevent MAC-hs buffer overflow. Hence, in addition to TSP queuing, D-TSP incorporates a credit-based flow control mechanism, necessitating the additional thresholds, L and H besides the TSP threshold R. Thus the overall D-TSP (logical) queue incorporates three thresholds as shown in Figure 5.

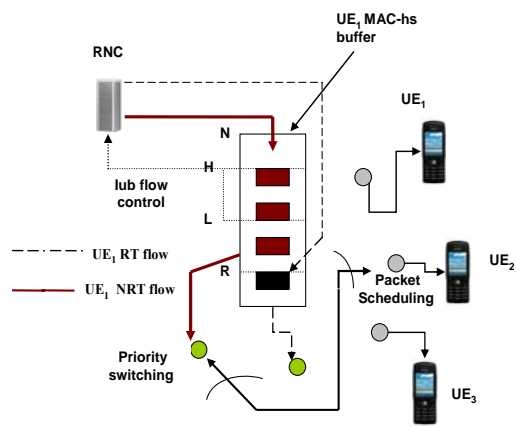

**Figure 5: D-TSP buffer management scheme shown for $UE_1$ RT and NRT flow queuing, priority handling and QoS control in HSDPA Node B MAC-hs.**

The D-TSP flow control mechanism issues credits which gives the number of data units of each flow to be transmitted from the RNC. Furthermore, the D-TSP flow control mechanism is designed to react to variation of the UE channel conditions, as well as buffer occupancy in order to mitigate buffer overflow and ensure efficient radio resource utilization. In addition, the D-TSP scheme incorporates time or transmission priority switching between the RT and NRT flows. Basically, for a given transmission opportunity assigned to the UE by the packet scheduler, when the head-of-the-line (HOL) RT packet delay is below a given delay budget, transmission priority is switched to the NRT flow, otherwise it remains with the RT flow. Referring to Figure 5, The D-TSP algorithm is described with the following assumptions and notations:

- Assuming a total buffer allocation of N Protocol data units (PDUs) for a given UE with a multimedia connection/session in the Node B MAC-hs in the HSDPA cell. Let R denote the total number of allowed RT PDUs in the UE's MAC-hs buffer.

- Let r(t) be the number of the UE's RT PDUs in the buffer at time t, while we denote the number of the UE's NRT PDUs at time t as n(t). From TSP principle, $0 < r(t) < R$ and $0 < n(t) < N$ at any given time t.

- Denote the lower D-TSP Iub flow control threshold as L, where $R < L$. Likewise the higher flow control threshold is given by H, where $L < H < N$.

- Let the nth user's buffer occupancy at time t be given by $q(t) = r(t) + n(t)$. The average buffer occupancy is estimated using a moving average filter with ith sample given by:

$$q_i = w \cdot q_{(i-1)} + (1-w) \cdot q(t) \qquad (1)$$

- Denote $\lambda_{rt}$ as the Guaranteed Bit Rate (GBR) of the RT flow (obtainable from bearer negotiation parameters).

- Let $\lambda'_{nrt}$ express the estimated average NRT flow data rate at the radio interface determined from:

$$\lambda'_{nrt\,(i)} = \alpha \cdot \lambda'_{nrt\,(i-1)} + (1-\alpha) \cdot \lambda_{nrt}(t) \qquad (2)$$

where $i$ is the ith TTI in which the user's NRT flow was transmitted during the UE's scheduling opportunity and $\lambda_{nrt}(t)$ is the amount of NRT data transmitted during the $i$th TTI. $\lambda_{nrt}(t) = 0$ if no NRT PDUs were transmitted in the $i$th TTI for the given UE.

- Let $k$ denote a parameter for buffer overflow control, while T denotes the inter-frame period for RNC-Node B frame transfer (10ms), and PDU_size, the MAC-d PDU size in bits.

- Assuming a given delay budget, DB for the UE RT PDU MAC-hs queuing. RT PDU inter-arrival time, $i$ at the Node B MAC-hs can be estimated from the already known GBR using:

$$i = \text{PDU\_size (bits)} / \lambda_{rt} \text{ (bits/sec)} \qquad (3)$$

- Thus, we can define a time priority switching control parameter $\delta$ given by: $\delta = DB / i$ (4)

- Assuming a discard timer (DT) is used to discard MAC-d PDUs of the UE's RT flow when MAC-hs queuing delay exceeding a given maximum delay budget $DB_{max}$. If Y is the maximum allowable downlink delay then:

$$DB_{max} = Y - (\text{external network delays + Core Network delays + Iub transfer delay}) \qquad (5)$$

With the above given assumptions and defined notations, the overall D-TSP scheme in MAC-hs operates as follows:

**Part 1: Credit allocation for multimedia UE:**

- Step 1: Compute per frame RT flow credit allocation
  $$C_{rt} = (\lambda_{rt} / \text{PDU\_size}) \cdot T \qquad (6)$$

- Step 2: Compute per frame maximum NRT credits

    $C_{NRTmax} = (\lambda'_{nrt} /PDU\_size) \cdot T \quad \text{if } q_i < L$
    $\quad\quad\quad\quad k \cdot (\lambda'_{nrt} /PDU\_size) \cdot T \quad \text{if } L \leq q_i \leq H$
    $\quad\quad\quad\quad 0 , \quad\quad\quad\quad\quad\quad\quad\quad \text{if } q_i > H \quad (7)$

- Step 3: Compute per frame NRT credit allocation

    $C_{NRT} = \min \{C_{NRTmax}, UBS_{NRT}\}$ where $UBS_{NRT}$ is the number of NRT PDUs present in the RNC for the UE. Hence, total per frame credit for the UE is $C_{rt} + C_{NRT}$.

**Part 2: TSP queue management for multimedia UE:**

- Step 1: For each arriving HS-DSCH data frame from RNC for the UE determine the flow class - RT or NRT.
- Step 2: If flow belongs to RT class, for each MAC-d PDU in the payload:

    If $(r(t) < R)$ *queue PDU at RT queue tail*
    Else *drop MAC-d PDU and update RT loss*
    Else If flow belongs to the NRT class, for each MAC-d PDU in the payload:

    If $r(t) + n(t) < N$ *queue PDU at buffer queue tail*
    Else *drop MAC-d PDU and update NRT loss*

**Part 3: Transmission priority control (D-TSP only):**

- For each scheduled UE transmission opportunity:

    IF $(r(t) < \delta$ AND RT HOL delay $< DB_{max}$ AND $n(t) > 0)$
    Time Priority = NRT flow
    Generate MAC-hs Transport Block from NRT PDUs
    ELSE
    Time Priority = RT flow
    Generate MAC-hs Transport Block from RT PDUs

### C. *Benefits of dynamic BMA for HSDPA multimedia traffic*

The application of a dynamic buffer management scheme -such as the described D-TSP- for QoS control at the wireless interface will yield the following benefits:

- Since NRT class uses RLC Acknowledged Mode for MAC-d PDU transfer, RLC layer performance improves dramatically due to reduced RLC round trip times and fewer RLC retransmissions. Even a small percentage loss of NRT PDUs is detrimental. In contrast, RT class uses RLC Unacknowledged Mode transfer where RT MAC PDU losses have no effect on RLC RTT.
- Improved TCP performance and application response times: Unlike the RT flow which will typically use UDP, the NRT flow in the HSDPA multimedia session utilizes TCP as the transport layer protocol. TCP offers reliability and flow control for NRT services and accounts for majority of data traffic in the Internet. TCP was originally designed for wired networks, but extensive research efforts have concentrated on optimizing TCP performance over wireless networks, including UMTS/HSDPA. Previous work such as [22], [23], [24], [25], have shown that starvation of TCP-based NRT flows in the lower layers have severe adverse effects on higher layer (RLC and TCP) protocol performance. Hence, D-TSP flow control and dynamic priority switching mechanisms can alleviate potential starvation of the NRT class with consequent TCP performance enhancement.
- Increased resource utilization efficiency: Loss of RT PDUs in the MAC-hs buffers incur no additional RAN or radio resources. On the other hand, NRT PDUs lost due to MAC-hs buffer overflow are retransmitted across all layers, thereby wasting network and transmission resources from end to end.

D-TSP, through its dynamic transmission priority control allows for trade-off of RT QoS within allowable constraints whilst maximizing NRT bandwidth allocation. In the next section D-TSP performance is investigated.

### III. D-TSP PERFORMANCE EVALUATION

In order to evaluate D-TSP for streaming RT traffic and TCP-based NRT traffic in a concurrent HSDPA user's multimedia session, we used the static equivalent, (s-TSP), and complete buffer sharing (CBS) as baseline schemes for comparison. The static TSP (s-TSP) scheme consists of the TSP queuing and flow control thresholds and mechanisms described in the previous section for the D-TSP scheme but without the dynamic transmission priority switching aspect (i.e. part 3). This means that static TSP always prioritizes the UE's RT packets (PDUs) for transmission.

With complete buffer sharing, NRT flow is guaranteed some bandwidth allocation at the radio interface in the presence of the RT streaming flow of the same user, because CBS inherently possesses some degree of buffer and transmission bandwidth allocation fairness [8]. For this reason, CBS provides a comparative baseline scheme to evaluate the NRT flow starvation mitigation capabilities of D-TSP.

Recall that D-TSP uses dynamic time priority switching in order to prioritize NRT transmission while RT flow delay is within a given delay budget. RT streaming being a *greedy source* traffic has the potential to cause NRT bandwidth starvation. So, the study is aimed at investigating the impact of RT streaming flow on the concurrent NRT TCP flow of the same HSDPA user, and the effectiveness of D-TSP in mitigating NRT flow starvation while still ensuring that RT streaming end-to-end QoS requirements are not violated.

### A. *Simulation model and configuration*

For the study, a HSDPA system model was developed with detailed UTRAN mechanisms including, RNC MAC queues, RLC layer AM and UM modes with ARQ retransmission for AM mode. In the Node-B, MAC-hs queues (applying CBS, s-TSP and D-TSP), HARQ processes, AMC schemes, and Packet Scheduling on the HSDPA air interface are modeled. Effect of the core network (CN) is abstracted as an assumed fixed delay to arriving packets. In the receiver (UE), we included SINR calculation and CQI reporting, HARQ processes, RLC modes with ARQ for AM retransmission, packet reassembly queues, peer TCP entity, and playout buffer for the streaming RT flow.

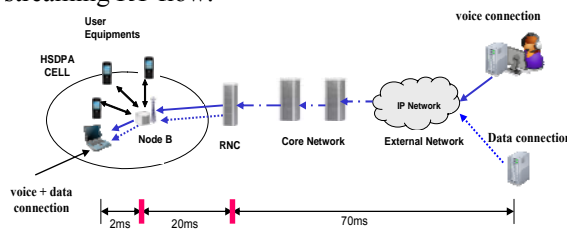

**Figure 6: Simulation model for D-TSP case study with RT streaming and TCP-based NRT flow in a concurrent UE session.**

In the experiments, a test user equipment ($UE_1$) is connected to the UTRAN and configured to receive 'multimedia' traffic of simultaneous 64 kbps Constant Bit Rate (CBR) encoded real-time video stream and non-real-time TCP streams in a simulated 120s streaming and file download session. The overall set up models a single HSDPA cell with fair time scheduling to *m* users as shown in Figure 6. A summary of the simulation parameters are given in Table 1.

Several scenarios with different HSDPA cell loads were considered i.e. *m* = 1, 5, 10, 20 and 30 users simultaneously active during the test UE's concurrent streaming and file download session of 120s duration.

**Table 1: Simulation parameters for D-TSP investigation.**

| Physical layer configuration | |
| --- | --- |
| HS-DSCH TTI | 2ms |
| HS-DSCH Spreading factor | 16 |
| HSDPA carrier frequency | 2GHz |
| Path loss Model | 148 + 40 log (R) dB |
| Transmit powers | Total Node-B power=15W, HS-DSCH power= 50% |
| Noise power | 1.214 e$^{-13}$ W |
| Shadow fading | Log-normal: σ = 8 dB |
| AMC schemes | QPSK ¼, QPSK ½, QPSK ¾, 16QAM ¼, 16QAM ½, 16QAM ¾ |
| No. of HS-DSCH codes | 5 |
| CQI letency | 3 TTIs (6ms) |
| No. of active HARQ processes | 4 |
| HARQ feedback latency | 5ms |
| MAC-hs configuration | |
| Packet Scheduling | Fair time scheduling |
| Iub flow control | Enabled |
| Parameters for flow control algorithm | α = 0.7; w= 0.7; k= 0.5 |
| D-TSP /s-TSP parameters | R = 32, L = 72, H = 144, N =192 |
| DB settings | DB = 40, 80, 120 and 160 ms |
| CBS parameter | N = 192 |
| Discard time DT timeout | 160ms |
| RLC configuration | |
| Operation mode (NRT flow) | Acknowledged |
| PDU delivery | In-Sequence |
| PDU size | 320 bits |
| RLC TX_window size | 1024 PDUs |
| RLC RX_window size | 1024 PDUs |
| SDU discard mode | After MaxDAT |
| MaxDAT | 6 attempts |
| Polling mechanism | RLC status every PDU |
| PDU retrans. delay | 200ms |
| Iub (RNC-Node-B) delay | 20ms |
| TCP configuration | |
| Maximum segment size | 512 bytes |
| RWIND | 32 KB |
| Initial CWIND | 1 MSS |
| Initial SS_threshold | RWIND |
| Fast retransmit duplicate ACKS | 3 |
| External + CN delays | 70ms |

*B. Buffer dimensioning*

Assuming a downlink maximum transfer delay of 250ms, the maximum MAC-hs delay budget $DB_{max}$ can be calculated from equation (5), given that assumed Core Network + external delay + Iub delay sum up to 90ms (see Table 1). It is reasonable to assume that RNC queuing contributes very little delay comparatively because the D-TSP flow control algorithm design ensures that RT PDUs are not held back in the RNC queues. Moreover, this was confirmed during the simulations. Thus, from equation (5) $DB_{max}$ = 160ms, and, therefore RT discard timer DT is set to 160ms.

For the 64 kbps CBR RT stream, $λ_{rt}$ = 64 kbps hence:
R = ($λ_{rt}$ * $DB_{max}$) /PDU_size = 32 PDUs

Likewise, assuming a maximum bit rate of 256 kbps for NRT flow and maximum average MAC-hs delay budget of 200ms:

Buffer size = (256 000 * 0.2)/PDU_size = 160.

Hence, we take a total buffer size N = 32 +160 = 192 PDUs in the MAC-hs for the UE. For the flow control thresholds we take H= 0.75* N = 144, and L =0.5 * H = 72.

*C. Performance metrics*

In the experiments, we consider MAC-hs queuing delay budgets, DB, of 40, 80, 120 and 160 ms which from equation (4), correspond to δ = 8, 16, 24 and 32 respectively. Note that the discard timer, DT discards RT streaming packets whose MAC-hs queuing delay ≥ 160ms from the head of the D-TSP/s-TSP queue. The performance metrics observed are:

- *Average end-to-end NRT throughput*: The time average of the throughput of the TCP-based NRT flow measured in the UE TCP layer.
- *RT PDU discard ratio*: The ratio of late RT streaming PDUs discarded in the MAC-hs as a result of DT timeout.
- *RT inter-packet playout delay*: The playout delay between successive packets of the RT streaming flow queued in the UE playout buffer after the first initial buffering delay of $DB_{max}$.

## IV. NUMERICAL RESULTS

*A. End-to-End NRT throughput evaluation*

Figure 7 plots the average end-to-end NRT flow throughput of an HSDPA end user (UE1) terminal running a session of simultaneous CBR encoded 64 Kbps RT video streaming and TCP-based file download. The average throughput is plotted against the number of users sharing the HSDPA channel in a single cell with fair time packet scheduling. The time average of the obtained throughput measured in the UE over a session period of 120s for δ = 8, 16, 24 and 32 delay budget settings are compared to that of s-TSP and CBS buffering. In all the scenarios UE1 is assumed to be located at 0.2 km from the base station and moving away at 3 km/h, while other users are placed at random positions in the cell. From Figure 7, we can see that in the single user scenario i.e. when UE1 occupies the channel alone, the D-TSP scheme in all DB settings give only slightly better throughput than s-TSP or CBS. This represents a lightly loaded HSDPA channel scenario where the user is being allocated all available channel codes in every TTI, the resulting high bandwidth allocation prevents NRT flow starvation despite the presence of the 'greedy source' RT streaming flow. For the same reason, increasing the D-TSP parameter does not yield any throughput improvement.

In the 5 user scenario ($UE_1$ sharing with 4 other UEs), it is interesting to note that at this point starvation of NRT flow starts to occur with s-TSP while CBS gives about 128 Kbps average throughput. Recall that s-TSP also incorporates an Iub flow control algorithm. In the experiments it was determined that the flow control algorithm effectively prevented buffer overflow, so no NRT PDUs were lost with s-TSP, indicating that starvation was the cause of end-to-end TCP throughput degradation. Since according to equation (7), the NRT flow credit allocation by the flow control algorithm depends on buffer occupancy, UE radio conditions and NRT flow throughput at the radio interface, the cause of NRT bandwidth starvation can only be attributed to the static prioritization of the greedy source RT streaming flow by s-TSP, which has no transmission switching mechanism. In the same 5 user scenario, it can be seen that D-TSP was effective in allowing NRT PDUs through by delaying the RT flow PDUs for up to the given delay budget. As D-TSP parameter increases (i.e. delay budget is relaxed more), a corresponding improvement in NRT throughput is noticeable. Also all the D-TSP configurations outperform CBS, indicating better fairness in bandwidth allocation to the NRT flow (a property which is inherent in the latter).

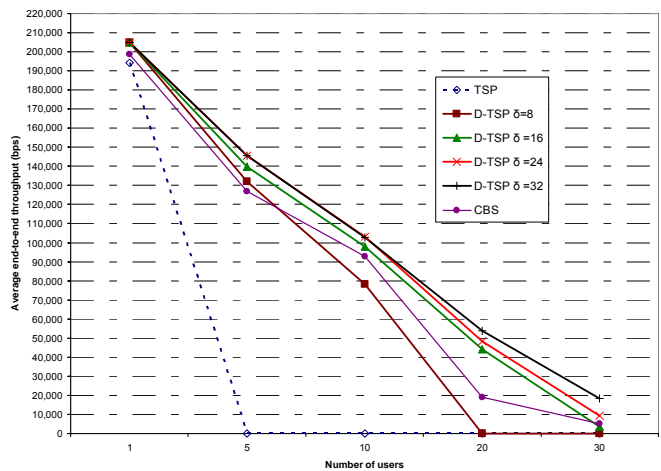

**Figure 7: End-to-end NRT throughput of UE1 for s-TSP, D-TSP ($\delta$ = 8, 16, 24, and 32 respectively) with 1, 5, 10, 20 and 30 active users in the HSDPA Cell.**

In the 10 user scenario, i.e. with heavier channel load, a similar trend is observed. D-TSP with DB of 80, 120 and 160 ms (i.e. $\delta$ = 16, 24 and 32) performed better than CBS. Again, starvation by RT streaming is apparent with s-TSP. In the 20 user scenario, again with heavier load and consequent less frequent scheduling opportunities, NRT flow starvation occurs with s-TSP and D-TSP $\delta$ = 8 (40ms delay budget). Only by increasing $\delta$ to 16 and above does D-TSP become effective in preventing NRT flow starvation and also exceeding CBS in average end-to-end throughput. With 30 users, we also encounter NRT flow starvation which is prevented again by D-TSP of delay budget 80ms and above. From the results, we conclude that D-TSP provides an effective mechanism through transmission priority switching to prevent imminent NRT flow starvation by a concurrent RT stream in a HSDPA user's session comprising both flows. Next we consider the impact of this mechanism on the streaming RT flow to see whether the trade-off for end-to-end NRT flow improvement was worthwhile.

*B. RT Streaming performance evaluation*

Since a discard timer DT is used to discard HOL packets with delay exceeding $DB_{max}$ (160ms), the RT streaming PDUs violating this bound will not be received at the $UE_1$. The D-TSP mechanism deliberately stalls RT PDUs to allow NRT PDUs transmission, thus increasing the probability of RT PDUs exceeding $DB_{max}$ and being discarded. Therefore in order to determine whether D-TSP provides the NRT end-to-end improvement without violating the RT streaming flow QoS bound, we recorded the number of RT PDUs discarded in the MAC-hs as a result of DT timeout. Figure 8 plots the RT PDU discard ratio vs. number of users in the cell. The plots are shown for D-TSP for $\delta$ = 24 and 32 corresponding to 120ms and 160ms delay budget. For s-TSP, D-TSP $\delta$ = 8 and 16, there were no RT PDUs discarded by the discard timer. Likewise for D-TSP $\delta$ = 24 single user, 5 user and 10 user scenarios; and also for D-TSP $\delta$ = 32 single user and 5 user scenarios.

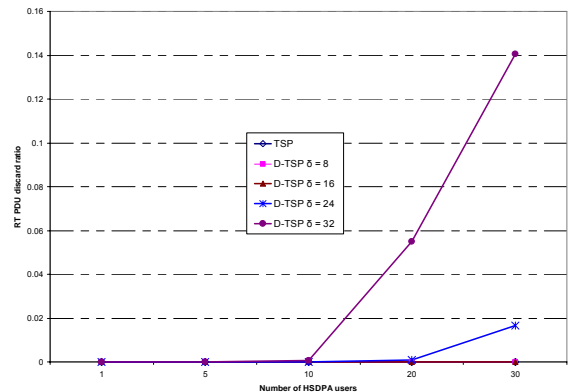

**Figure 8: RT PDU discard ratio vs. number of users.**

It is clear from the graph that with less than 30 users in the cell, D-TSP with 120ms delay budget can guarantee less than 2% discard with discard timer set to 160ms. However, when D-TSP is set with 160ms delay (which is the upper limit), with 20 users, about 5% of the PDUs are discarded by the DT, while with 30 users about 14 % of the RT PDUs are discarded by the DT. This implies that a 160ms delay budget setting for D-TSP is too high to be used in 30 user scenario without severely compromising the streaming RT flow QoS. However, it is worth noting also that the delay of RT PDUs is not due to D-TSP switching mechanism alone but also the high channel loading is a major contributing factor. Hence, from the results we conclude that considering both results (Figures 7 and 8 together) D-TSP is still effective in NRT throughput enhancement whilst keeping RT streaming losses to a minimum that will not violate its required QoS.

Lastly, we consider the $UE_1$ RT playout buffer to observe the effect of D-TSP on the end-to-end performance of the RT streaming flow. Since the RT streaming video is assumed to be 64 kbps CBR encoded, the arriving packets were buffered and played out at 64 kbps after an initial buffering delay equal to the maximum MAC-hs delay budget $DB_{max}$ of 160ms. We measured the inter-packet playout delay i.e. the delay between each successive packet played out from the UE buffer. A constant delay is expected if the buffer always contains a packet for playout, otherwise if the buffer empties at certain times, delay spikes will occur. Figure 9 shows the observed delays between successive played out streaming packets for s-TSP over the 120s session with simultaneous NRT and RT streaming flows for all the channel load scenarios. The inter packet delay is observed to be constant at 0.005s (corresponding to 64 kbps playout rate of 320 bit long packets) indicating no playout jitter. Hence the de-jittering buffer was effective in eliminating any jitter in the arriving packets. The same result was obtained for D-TSP $\delta = 8$ and 16 (not shown).

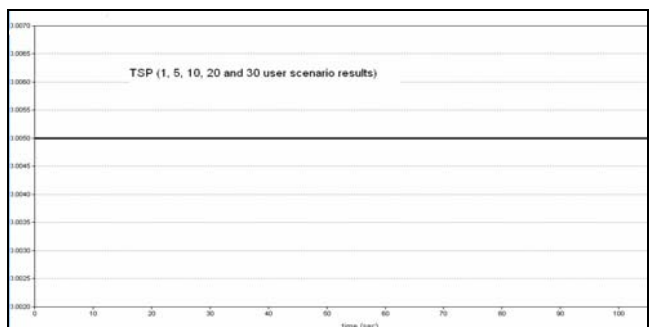

**Figure 9: RT inter-packet playout delay for all s-TSP scenarios. The same constant playout rate was obtained for all user scenarios of D-TSP $\delta= 8$ and D-TSP $\delta= 16$.**

Figure 10 shows the results for D-TSP 120ms ($\delta= 24$) for 20 user and 30 user scenario, while Figure 11 shows that of D-TSP 160 ms ($\delta= 32$) for 10, 20 and 30 user scenarios. All the omitted results for these D-TSP settings showed constant playout rate as in Figure 9. The 20 user scenario in Figure 10 (top half) showed only a few instances where the delay spiked (i.e. playout gaps in successive packets). This delay spikes correspond to periods of empty buffer and being few, we can assume minimal impact on the RT stream quality. The same goes for the 10 user scenario of D-TSP 160ms in Figure 11. Whereas, for the D-TSP 120ms 30 user scenario in Figure 10 and D-TSP 160ms 20 and 30 user scenarios in Figure 11, the delay spikes are more frequent depicting high playout jitter which will severely compromise playout quality. On the other hand, we note that degradation in RT streaming QoS in the UE in these scenarios cannot be attributed to the effect of D-TSP alone, but also to channel congestion due to higher cell loading. Nevertheless, the results prove that with careful parameter setting, D-TSP can operate within end-to-end RT streaming QoS constraints.

## V. CONCLUDING REMARKS

The dynamic buffer management scheme presented and evaluated in this paper allows effective QoS management of multiple flows in an end-user multimedia session with concurrent RT and NRT flows in 3.5G wireless networks and similar beyond 3G systems with buffering at the air interface. The novelty of the scheme lies in not only utilizing time and space priorities in a combined manner to suit the different QoS requirements of the RT and NRT flows, but also in employing transmission priority switching to further optimize QoS control. Based on evaluations in a simulated HSDPA system, the proposed buffer management scheme was efficient in mitigating end-to-end NRT bandwidth starvation whilst simultaneously maintaining acceptable RT streaming flow QoS for the UE multimedia session comprising both flows.

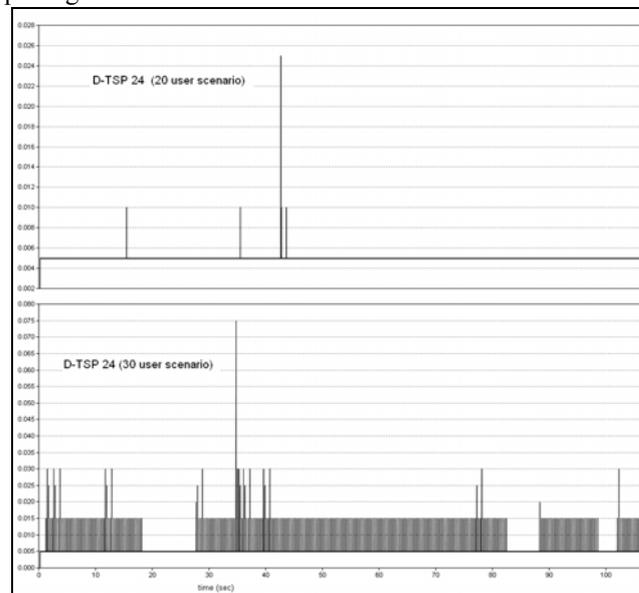

**Figure 10: RT inter-packet playout delay for D-TSP $\delta= 24$. 20 and 30 user scenarios are shown. The scenarios with fewer users gave constant inter-packet playout delay as in Figure 6.**

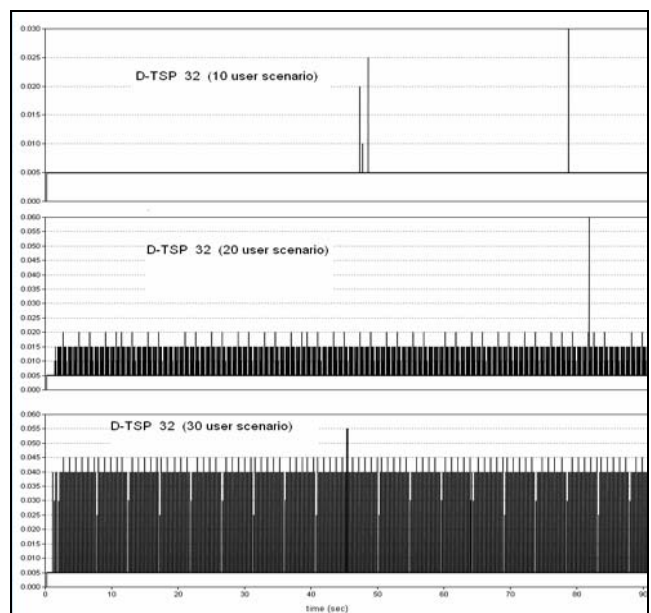

**Figure 11: RT inter- packet playout delay for D-TSP $\delta= 32$. 10, 20 and 30 user scenarios are shown. Scenarios with fewer users gave constant inter-packet playout delay as in Figure 6.**

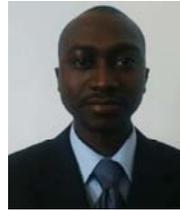

Suleiman Y. Yerima is with the Integrated Communications Research Centre (ICRC), University of Glamorgan, UK. He was awarded a Ph.D. from the University of Glamorgan, UK in July 2009 for his thesis titled 'Quality of Service Optimization of Multimedia Traffic in Mobile Networks'. He received his MSc degree (with Distinction) in Personal, Mobile and Satellite Communications from the University of Bradford, UK in 2004 and holds a first class honours degree in Electrical and Computer Engineering from Federal University of Technology, Minna, Nigeria. In 2006, he was awarded the prestigious Higher Education Funding Council for Wales (HFCW) ORSAS research scholarship.

Yerima was the recipient of the 2009 World Congress on Engineering (WCE 2009) Best Student Paper Award of the International Conference of Wireless Networks held in July. His current research interests are in resource management, Quality of Service, modelling and performance evaluation of mobile/wireless networks. He has reviewed, authored and co-authored several papers in his research area of interest, and has contributed to articles and book chapters in the area of HSDPA technology.

Khalid Al-Begain is a Professor of Mobile Computing and Networking and the Director of the Integrated Communications Research Centre (ICRC) at the University of Glamorgan, UK. He received his Ph.D. in Communications Engineering in 1989 from Technical University of Budapest, Hungary. He has been working in different universities and research centres in Jordan, Hungary, Germany and the UK. He has led and is leading several projects in mobile Computing, wireless networking and performance evaluation. He is the President of the European Council for Modelling and Simulation, UNESCO Expert in networking, Senior Member of the IEEE and IEEE Communications and Computer Societies, British Computer Society Fellow and Chartered IT Professional, member of the UK EPSRC College (2006–2009), Wales Representative to the IEEE UK & RI Computer Chapter MC, and UK representative to the COST290 Action management committee.

He has lead a number of projects including EPSRC funded on multimedia multicasting and EU FP6.

Al-Begain has been the General Chair of ten international conferences including three in 2008. In particular, he is the founder and general chair of the IEEE International Conference and Exhibition on Next Generation Mobile Applications, Services and Technologies with the next to be held in Cardiff in September 2009. He edited 12 books, co-authored one book and more than 150 papers in refereed journals and conferences. He is currently leading the project on the IMS based next Generation service and applications creation environment at ICRC supported by major industries. This facility is one of very few in universities worldwide.